\documentclass[10pt]{article}

\def\final{0}

\usepackage{csc}
\usepackage{algpseudocode}
\usepackage{graphicx}
\usepackage{forest}
\usepackage{amsmath}
\usepackage{amssymb}
\usepackage{fullpage}
\usepackage{comment}
\usepackage{commath}
\usepackage{amsthm}
\usepackage{setspace}
\usepackage[margin=1in]{geometry}
\usepackage{float}
\usepackage{framed}
\usepackage{algorithm}
\usepackage{tikz}
\usepackage[backref]{hyperref}
    \definecolor{darkcolor1}{RGB}{15, 43, 109}
    \definecolor{darkcolor2}{RGB}{90, 15, 28}
    \hypersetup{
    colorlinks=true,
    urlcolor=blue,
    linkcolor=darkcolor1,
    citecolor=darkcolor1,
    linktocpage=true
    }
\usepackage{microtype}

\excludecomment{question}

\def \E {{\mathbb{E}}}

\newcommand{\iprod}[2]{\langle #1, #2 \rangle}

\ifnum\final=0
\newcommand{\mynote}[2]{\marginpar{\color{#1}\tiny\sf #2}}
\newcommand{\mybignote}[2]{{\color{#1} $\langle \langle$ #2$\rangle \rangle$}}
\else
\newcommand{\mynote}[2]{}
\newcommand{\mybignote}[2]{}
\fi

\newcommand{\eps}{\varepsilon}
\DeclareMathOperator{\polylog}{polylog}

\newcommand{\ex}[2]{{\ifx&#1& \mathbb{E} \else \underset{#1}{\mathbb{E}} \fi \left[#2\right]}}
\newcommand{\pr}[2]{{\ifx&#1& \mathbb{P} \else \underset{#1}{\mathbb{P}} \fi \left[#2\right]}}
\newcommand{\var}[2]{{\ifx&#1& \mathsf{Var} \else \underset{#1}{\mathsf{Var}} \fi \left[#2\right]}}

\newcommand{\nope}[1]{}

\renewcommand{\epsilon}{\varepsilon}

\newcommand{\prot}{\rightleftharpoons}

\newtheorem{thm}{Theorem}[section]
\newtheorem{defn}[thm]{Definition}
\newtheorem{cor}[thm]{Corollary}

\newtheorem{rem}[thm]{Remark}

\newtheorem{theorem}[thm]{Theorem}
\newtheorem{corollary}[thm]{Corollary}
\newtheorem{lemma}[thm]{Lemma}

\newtheorem{proposition}[thm]{Proposition}
\newtheorem{definition}[thm]{Definition}
\newtheorem{claim}[thm]{Claim}

\newtheorem*{upperboundmain}{Theorem~\ref{thm:upper_bound_main}}
\newtheorem*{lowerboundmain}{Theorem~\ref{thm:lower_bound_main}}
\newtheorem*{sdpappendix}{\hypertarget{app:prop:SDP}{Proposition~\ref{prop:SDP}}}
\newtheorem*{bdktappendix}{\hypertarget{app:lem:bdkt}{Lemma~\ref{lem:bdkt}}}
\newtheorem*{weightedappendix}{\hypertarget{app:lem:weighted_target_charging}{Lemma~\ref{lem:weighted_target_charging}}}

\theoremstyle{definition}

\newcommand{\todo}[2]{}
\newcommand{\sasho}[1]{}
\newcommand{\haohua}[1]{}

\let\originalleft\left
\let\originalright\right
\renewcommand{\left}{\mathopen{}\mathclose\bgroup\originalleft}
\renewcommand{\right}{\aftergroup\egroup\originalright}

\title{Private Online Release of Adaptive Linear Queries}
\author{Aleksandar Nikolov\thanks{Department of Computer Science, University of Toronto. Supported in part by an NSERC Discovery Grant (RGPIN-2021-03206), and the Canada Research Chairs program (CRC-2020-00004).} \and Haohua Tang\thanks{Department of Computer Science, University of Toronto. Supported in part by an NSERC Discovery Grant (RGPIN-2021-03206), and the Canada Research Chairs program (CRC-2020-00004).}}
\date{}
\begin{document}
\maketitle
\begin{abstract}
    In this paper we study answering counting and other linear queries online under differential privacy constraints. Recent work by Nikolov, Tang, and Ullman [STOC 2026] gave the first online algorithm that answers \emph{obliviously chosen} online linear queries privately, and achieves poly-logarithmic competitiveness against an optimal offline factorization mechanism. This also implied poly-logarithmic competitiveness against arbitrary private offline mechanisms when the private dataset is large enough. The error guarantees of their algorithm, however, can fail when the queries are adaptive.  We give a new algorithm that also achieves poly-logarithmic competitiveness, while being able to handle \emph{adaptively chosen} queries. This implies that there is no large separation between the optimal error bounds for privately answering, respectively, offline linear queries, obliviously chosen online linear queries, and adaptively chosen online linear queries. Our algorithm combines the online factorization approach from prior work with a novel layered Euclidean iterative database construction, presenting a new way to use factorization in private algorithms.
    
    We also give a lower bound that shows that some poly-logarithmic factor in the competitive ratio is unavoidable. Our lower bound first shows that any interactive private mechanism implies a deterministic online factorization algorithm, and then gives a new lower bound for the competitive ratio of any deterministic online factorization algorithm.
\end{abstract}
\vfill\newpage
\vfill\newpage
\section{Introduction}
In the 20 years since it was introduced, differential privacy \cite{DworkMNS06} has become the golden standard for analyzing data with rigorous individual privacy guarantees. Roughly, differential privacy requires that the output of the data analysis mechanism on a pair of neighboring datasets be similar. In this paper, as is standard, we assume that a dataset is a sequence of data points from some finite universe $\mathcal{U}$, and
say $X$ and $X'$ are neighboring if one can obtain $X'$ from $X$ by adding or removing one data point. This is the add/remove, or unbounded notion of neighborhood. The formal definition of differential privacy is as follows.
\begin{definition}[Differential privacy \cite{DworkMNS06}]
A randomized mechanism $\mathcal{M}$ is $(\varepsilon,\delta)$-differentially
private if, for any pair of neighboring datasets $X,X'$, and every measurable set $S$ of outputs,
\[
    \Pr[\mathcal{M}(X)\in S]
    \le e^\varepsilon \Pr[\mathcal{M}(X')\in S]+\delta.
\]
\end{definition}

Absent privacy constraints, data analysts often interact repeatedly with their data, by first doing exploratory analysis, before settling on a final collection of statistical analyses. 
Differential privacy doesn't allow direct data access, and, instead, the analyst must interact with the data via the private mechanism. To offer an interactive workflow similar to the non-private setting, the mechanism should itself be interactive, and be able to answer the analyst's queries immediately, before new queries are specified based on the answers to the initial ones. 
This motivates the online/interactive private query release problem: design mechanisms that receive a new query at each time step, output a reasonably accurate answer to it, and ensure that the entire interaction transcript is differentially private. 

In the interactive query release problem, the queries could be \emph{obliviously} chosen, i.e., unknown to the mechanism but fixed ahead of time, or \emph{adaptive}, where new queries depend on the transcript so far.
Understanding the adaptive queries case is of particular practical importance. 
For example, suppose we have a dataset of demographic statistics. 
A data analyst may first want to understand correlations between ethnicity and a certain income level, and only after seeing a high correlation with one of the ethnicities, they ask more queries that may reveal confounding factors. These queries are adaptively chosen, so the data releasing mechanism must have both privacy and accuracy guarantees against adaptively chosen queries. 
Another advantage of interactive mechanisms that can accurately answer adaptively chosen queries is that other algorithms can use them as  subroutines. As a toy example, suppose we have a dataset of ages of a group of people, and the type of query allowed is to count the number of people  inside an age range. We can then use binary search to find an approximate median age for the group using adaptively chosen range queries. Adaptivity is important since binary search chooses its next query based on answers to the previous ones.

Overall, these examples show the importance of adaptiveness of queries for private interactive data release.

Alongside the motivation from private data analysis, asking adaptive queries poses statistical validity questions even when privacy is not a concern. In particular, adaptively analyzing a fixed dataset may lead to results that overfit to the data and do not generalize to the underlying population~\cite{GelmanL13}. Prior work~\cite{DFHPRR15,BassilyNSSSU16} has shown that differential privacy can be used as a tool to ensure validity of adaptive data analysis: an analyst that interacts with the data via a differentially private mechanism, even adaptively, cannot overfit to the dataset. Slightly more precisely, if an interactive mechanism answers adaptively chosen queries privately, and accurately with respect to the dataset, then its answers are also accurate with respect to the population. This provides further motivation for the interactive private query release problem in the adaptive setting. \sasho{maybe check this with the machine to see if we are missing citations. i have not paid attention to this line of work since 2018 or so}

Despite the substantial motivation for developing interactive private mechanisms that can accurately answer queries posed by an adaptive analyst, optimal mechanisms for this problem are poorly understood, even for the fundamental example of counting queries, i.e., queries of the type, ``How many data points satisfy a given property?'' By contrast, there is a substantial body of work in differential privacy that has identified approximately optimal mechanisms for answering offline counting queries, and, more generally, linear queries (see the definition below)~\cite{HardtT10,BDKT12,NikolovTZ13,BlasiokBNS19,EdmondsNU20}. These mechanisms need all queries to be specified ahead of time, run in time polynomial in the dataset size, universe size, and number of queries, and achieve error within poly-logarithmic factors of the optimal error achievable by any differentially private mechanism.  Our goal in this paper is to achieve similar guarantees in the interactive setting with adaptive queries. 

\subsection*{Linear Queries}

We study private algorithms that take as input a dataset $X\in \mathcal{U}^n$, where $n\in \N$ and the data universe $\mathcal{U}$ has finite size $N$. Note that we adopt the add/remove definition of neighboring datasets, so the dataset size $n$ is not public. We consider linear queries, defined below, which characterize a broad class of queries of interest in applications, including all counting queries. 
\begin{definition}[Linear Query]
    A linear query is specified by
a function $q:\mathcal{U}\to\mathbb{R}$. For a dataset
$X=(x_1,\ldots,x_n)\in\mathcal{U}^n$, its answer is
\[
    q(X) \coloneq  \sum_{i=1}^{n} q(x_i).
\]
\end{definition}
Counting queries are the special case when $q$ only takes values $0$ or $1$. Linear queries have been studied extensively under differential privacy, and many real world data releases, such as the 2020 US Census~\cite{CensusDP17}, are of this form. Even more restricted classes of counting queries, like marginal queries~\cite{BarakCDKMT07,LebedaNT26, XiaoHTDZK26}, have received significant attention.

It is convenient to represent the dataset as a vector, known in the literature as the histogram representation. 
\begin{definition}[Histogram]
    The histogram of a dataset $X \in \mathcal{U}^*$ is the vector
$h\in\mathbb{Z}_{\ge 0}^{\mathcal{U}}$, where $h_x$ counts the number of
occurrences of $x$ in $X$, i.e., $h_x \coloneq  |\{i: x_i = x\}|$ for $X = (x_1, \ldots, x_n)$. 
\end{definition}
A linear query can then be viewed as the inner product of a query row vector $q_t$ and the histogram $h$ corresponding to the dataset, and the answer to a workload of linear queries can be written as a matrix-vector product of a matrix representing the queries and the histogram, as shown below.
\begin{definition}[Query Matrix]
    For a workload of linear queries $q_1,\ldots,q_m:\mathcal{U}\to\mathbb{R}$, the
query matrix $Q\in\mathbb{R}^{m\times\mathcal{U}}$ of the workload is defined by
$Q_{i,u}=q_i(u)$. The answer to the workload is \[Q(X)\coloneq \begin{bmatrix}q_1(X)\\\ldots\\q_m(X)\end{bmatrix}=Qh.\]
\end{definition}

Note that, for any two neighboring datasets $X$ and $X'$, the corresponding histograms satisfy $\|h-h'\|_1 \le 1$. This easily implies that $\|Q(X) - Q(X')\|_2 \le \|Q\|_{1\to 2}$, where $\|Q\|_{1\to 2}$ is the largest $\ell_2$ norm of a column of $Q$. I.e., the $\ell_2$-sensitivity of $Q$ is $\|Q\|_{1\to 2}$.

An important, and widely used, class of private mechanisms for releasing answers to workloads of linear queries are noise-adding, also known as oblivious, mechanisms. These mechanisms output $Q(X) + Z$ where $Z$ is a noise vector from some distribution that depends on $Q$ but not  on $X$. A simple example of such a mechanism is the Gaussian noise mechanism \cite{DworkR14}, in which $Z$ is a vector of independent and identically distributed (i.i.d.) Gaussians, with mean $0$, and standard deviation scaled to the $\ell_2$-sensitivity $\|Q\|_{1\to 2}$ of $Q$. Both the privacy and accuracy guarantees of this mechanism extend, in a straightforward way, to the interactive setting with adaptively chosen queries, as long as the $\ell_2$-sensitivity of the queries is bounded: one just needs to add the Gaussian noise to each query answer ``on the fly'' as the queries arrive. 

\subsection*{(Online) Factorization Mechanisms}
Factorization mechanisms~\cite{LiMHMR15,EdmondsNU20} provide a principled way to achieve better error than the i.i.d.~Gaussian noise by adding \emph{correlated} Gaussian noise. Given the query matrix $Q$, one can find a factorization $Q=LR$, then answer $Qh+LZ$, where $Z$ is again a vector of i.i.d.~Gaussians, this time with standard deviation scaled proportionally to  $\|R\|_{1\to 2}$. The maximum noise variance over queries in the workload achieved by this mechanism is proportional to $\|L\|_{2\to\infty} \|R\|_{1\to 2}$, where $\|L\|_{2\to\infty}$ is the maximum $\ell_2$ norm of a row of $L$. Optimizing the error of a factorization mechanism then corresponds to minimizing  $\|L\|_{2\to\infty} \|R\|_{1\to 2}$ over all choices of $L$ and $R$ that give $Q = LR$. The minimum such value is denoted $\gamma_2(Q)$, the $\gamma_2$ norm of $Q$. Together with an optimal factorization, it can be computed efficiently using convex optimization. Factorization mechanisms generalize and allow us to optimize widely used mechanisms such as the binary tree mechanism~\cite{DworkNPR10,ChanSS11}. It was shown by Edmonds et al.~\cite{EdmondsNU20} that, in the offline setting, the optimal factorization mechanism is also asymptotically optimal among all noise-adding mechanisms, and, when the dataset is large enough, among all differentially private mechanisms.

Intuitively one might expect that factorization mechanisms are restricted to the offline setting, since one needs to know the whole query matrix $Q$ in order to find its optimal factorization $Q = LR$. Counter to this intuition, it was shown in recent work by Nikolov, Tang, and Ullman~\cite{NTU26} that an approximately optimal factorization can be computed when the rows of the matrix $Q$ arrive online. The value of their online factorization is within a poly-logarithmic factor from the optimal offline factorization. Nikolov et al.~then also showed that this online factorization can be used to answer online queries privately, similarly to how an offline factorization is used. Their mechanism is still private against adaptive queries, but its error bound breaks against carefully chosen adaptive queries. This motivates us to ask: Is there a mechanism that works for adaptively chosen queries with a similar accuracy guarantee, and, conversely, can one show a lower bound on the competitive ratio, i.e. can one actually do as well as the offline case without even losing a log factor? In this paper, we address these two problems.

Let us describe the online factorization framework of Nikolov, Tang and Ullman~\cite{NTU26}. 
\begin{definition}[Online Factorization Problem]
    Let $Q \in \mathbb{R}^{m \times N}$ have rows
$q_1,\ldots,q_m$, revealed sequentially, and let $Q_t$
consist of its first $t$ rows. The number $m$ is not
known in advance.

Starting from the empty matrix $R_0$, an online
factorization algorithm processes each arriving row $q_t$
by appending additional rows to $R_{t-1}$ to obtain $R_t$,
and producing a row vector $\ell_t$ such that
\[
    \ell_t R_t = q_t, \qquad \|R_t\|_{1\to 2} \le 1.
\]
Let $L_t$ contain the rows $\ell_1,\ldots,\ell_t$, padded with zeros as necessary. The algorithm achieves factorization value $\gamma$ on $Q$ if
\[
    \|L_m\|_{2\to\infty} \le \gamma.
\]
\end{definition}

Nikolov, Tang, and Ullman showed that there exists an online factorization algorithm that is polylogarithmically competitive against the best offline factorization at every time step.
\begin{lemma}[\cite{NTU26}, Theorem~1.4]\label{lem:online_factorization_ntu}
    Fix $c>0$. There exists an online factorization
algorithm whose factorization value on every matrix
$Q \in \mathbb{R}^{m\times N}$ satisfies
\[
    \gamma
    =
    O\!\left(
        \gamma_2(Q)\,
        \left[
            \log\!\left(
                \frac{\gamma_2(Q)}{\|q_1\|_\infty}
            \right)
        \right]^{\frac12+c}
        (\log N)^3
    \right).
\]
\end{lemma}
This algorithm first solves a semidefinite program online to obtain a factorization in which the column norms of $R_t$ are bounded only on average, and then uses some classical combinatorial tricks to satisfy the stronger requirements of online factorization. In our paper, we use this algorithm as a black box.

It is straightforward to use the online factorization to privately answer online linear queries. When a new query arrives, we use the online factorization to compute $\ell_t$ and $R_t$. We compute $R_t h + Z_t$, where $Z_t$ is a vector of i.i.d.~Gaussian noise variables, one for each row of $R_t$. We can do so by just computing answers for the new rows in $R_t$ and adding Gaussian noise to them. Then we take an inner product of $\ell_t$ and $R_t h + Z_t$ to compute an answer to $q_t$. The privacy of this mechanism holds against adaptive queries, and follows from the privacy of the Gaussian noise mechanism. It is also straightforward to prove accuracy guarantees against oblivious queries. When the queries are adaptive, however, the analyst may choose $q_t$ to correlate with the noise $Z_t$, achieving much larger error than in the oblivious setting. Thus, we need a novel way of using the online factorization when the queries are adaptive.

\subsection*{Our Results}
In this subsection we introduce our results. We defer the formal definition of interactive algorithms, their privacy and accuracy, to Section~\ref{sect:prelims} - roughly, the algorithm receives query $q_t$ at time step $t$, outputs $o_t$ before receiving the next query, while keeping the entire transcript private, and every answer $o_t$ accurate.

Our first main result is a new algorithm that privately answers adaptively chosen linear queries with polylogarithmic competitiveness against the best offline factorization mechanism. Compared with the aforementioned algorithm in \cite{NTU26}, our algorithm has slightly worse log factors, but maintains both privacy and accuracy under adaptive queries.
\begin{theorem}[Main Upper Bound, Informal, {\hyperref[thm:main_UB_full]{\textcolor{blue}{Full Version in Section~\ref{sect:upper_bound}}}}]\label{thm:upper_bound_main}
    There exists a differentially private mechanism that w.h.p. answers adaptively chosen linear queries $Q=(q_1,...,q_m)$ over any dataset $X\in \mathcal{U}^n$ with error
    \[\max_{t\le m}
        |o_t-q_t(X)|= O(\gamma_2(Q)\polylog(N))\]
        where $o_t$ is the mechanism's output at time $t$.
\end{theorem}
In a previous work~\cite{BunSU17}, Bun, Steinke, and Ullman posed the open problem of whether there is a polynomial separation between adaptive and oblivious analysts for privately answering linear queries online. The work by \cite{EdmondsNU20} already shows that even for the offline case, when the dataset is sufficiently large, the best error one can achieve is in $\Omega(\gamma_2(Q))$. Our new upper bound matches this lower bound up to poly-logarithmic factors, giving an answer to the open problem for large datasets, and showing that, in this regime, there is only poly-logarithmic separation between offline query release, online oblivious query release, and online adaptive query release.

Our upper bound's algorithm uses three key ingredients: the online factorization algorithm from \cite{NTU26}, a layered version of the Euclidean Iterative Database Construction (IDC) algorithm from \cite{GRU12} and an extended version of the target charging analysis from \cite{CohenL23}.  We use factorization to get $\ell_tR_t=q_t$ at each time step $t$. The online factorization framework ensures that $\norm{\ell_t}_2\le \gamma$, and for any pair of neighbouring $h,h'$ we have $\norm{R_th-R_th'}_2\le 1$. One can view $\ell_t$ as a query, and $R_th$ as a private point $x$. Then our problem translates to a problem where we are given a point $x$ as the private input, neighboring inputs differ by at most $1$ in $\ell_2$ distance, and queries ask for an inner product of $x$ with a vector of $\ell_2$ norm at most $\gamma$. We use a modified version of IDC to answer these queries. 

The original Euclidean IDC algorithm keeps a public dataset $\hat{x}$  and tries to answer each query with it. Upon receiving query $q$ it tries to answer it with $\iprod{q}{x}$, and uses a private test like $\operatorname{AboveThreshold}$ to determine if the error would be too large. If the test returns positive, it first releases a new noisy answer $z$ to the query, then updates the public dataset by shifting a fixed amount toward the direction of $q$. With some minor modification, we can use this method to achieve a competitive ratio of $\tilde{O}(N^{1/4})$. Instead of using the Euclidean IDC as a blackbox, we introduce a layered structure: we maintain a partition of $\R^d$, where $d$ is the dimension of $x$; each layer corresponds to a pairwise orthogonal subspace. Within each layer, we run a classical Euclidean IDC for queries and the public dataset projected onto the corresponding subspace. The update rule we use is slightly different: first release a noisy estimate, and shift the public dataset along the direction of the query not by a fixed amount, but to match this estimate. The key novelty here in our algorithm is that, whenever an update is triggered, we ``kick out'' the span of the update direction to the next layer, where, at the beginning, the first layer contains the entire space $\R^d$. This layered design will make sure the dimension of the subspaces ever handled by each layer decreases geometrically, thus bounding the total number of layers. This geometric decrease is derived by a potential argument inspired by the one used for the original Euclidean IDC. Our potential decomposes the original IDC potential $\|\hat{x} - x\|_2^2$ according to the layers - within each layer, project $\hat{x} - x$ onto the layer's subspace and then take the $\ell_2$ norm squared. At a high level, this layered structure allows us to make further updates inside each layer without ``damaging what has been fixed'', and achieve a logarithmic competitive ratio. 

The target charging analysis by Cohen and Lyu~\cite{CohenL23}, at a high level, allows a tighter composition of a sequence of private mechanisms by bounding the overall privacy cost using a count of occurrences of some carefully chosen outputs. This chosen set of outputs is called a ``target''. The advantage of the technique is that private mechanisms in the composition that do not produce a target output do not incur privacy loss. Unlike the Cohen and Lyu analysis, for our algorithm we need to allow the privacy parameter for each individual algorithm to be different, and to be chosen adaptively based on the output of earlier algorithms. To be more specific, the algorithms we are composing are the private tests called by the layered IDC algorithm, where the ``target'' would be a positive test, i.e. large error detected. Here the layered structure and kicking out directions is again crucial: since, effectively, only positive tests incur a privacy loss, and the direction used in any positive test is kicked out of the layer, a privacy loss will not be incurred in this direction again in the same layer. This allows us to bound the overall privacy loss in each layer in terms of the $\ell_2$ sensitivity of $x$, which is just $1$. By contrast for standard IDCs the privacy loss grows with the number of updates.

We also give a new lower bound showing that, compared with the offline factorization, a competitive ratio of about the square root of the log of the universe size is necessary. This is an online extension of the lower bound by \cite{EdmondsNU20}. 
\begin{theorem}[Main Lower Bound, Informal, {\hyperref[thm:main_LB_full]{\textcolor{blue}{Full Version in Section~\ref{sect:lower_bound}}}}]\label{thm:lower_bound_main}
    For sufficiently large $n$, any private mechanism that answers online queries needs error of 
    \[\Omega\left(\gamma_2(Q)\frac{\sqrt{\log N}}{\log\log N}\right)\]
    on some obliviously chosen sequence of queries $Q$.
\end{theorem}

Our lower bound shows that even if the queries are obliviously chosen, some log factor is necessary in the competitive ratio. The argument of our lower bound consists of two parts: the first half is essentially the same as in the offline version - in the offline world, \cite[Theorem~30]{EdmondsNU20} shows that a private algorithm implies a factorization of the same query matrix, and we show that a private online algorithm implies an online factorization algorithm of the same query matrix too. The second part is to show a lower bound for online factorizations, i.e. showing every online factorization algorithm's competitive ratio is lower bounded on a family of linear queries. In particular, inspired by \cite{BunSU17}, we choose to use a special set of threshold queries that forms a sequence of binary search steps, and we show that no online factorization algorithm can achieve a very small value on every single search path. This argument also gives a lower bound of \(\Omega(\frac{\sqrt{\log N}}{\log \log N})\) for the worst case competitive ratio for any deterministic online factorization algorithm. 

\subsection*{Related Works}

As mentioned above, the problem of designing simple and efficient algorithms that release answers to offline linear queries with nearly optimal error is the subject of a long line of work. The best known bounds for large datasets and $(\varepsilon, \delta)$-differential privacy are achieved by factorization mechanisms~\cite{EdmondsNU20}.

Several interactive mechanisms have been proposed for releasing answers to adaptive linear queries: the Median Mechanism~\cite{RothR10}, and the Private Multiplicative Weights Mechanism~\cite{HardtR10} are two examples, where the latter mechanism also achieves optimal error bounds for \emph{worst case} counting query workloads~\cite{BunUV14}. Both of these mechanisms are generalized by the Iterative Database Construction (IDC) framework~\cite{GRU12}, which we already mentioned above.



A systematic study of the cost of privately releasing answers to online queries was initiated by Bun, Steinke, and Ullman~\cite{BunSU17}. Among other questions, they investigated possible separations between offline, oblivious, and adaptive queries. They showed an exponential separation between the oblivious and adaptive settings for a specially designed class of \emph{non-linear} queries. They left open the question whether a similar separation exists for a more natural class of queries, such as workloads of linear queries. Our work answers this question and shows that for linear queries all separations are at most poly-logarithmic. They also gave an error upper bound for answering $m$ adaptively chosen online threshold queries of $O(\log m+\polylog n)$, where the dataset points are from the infinite universe $[0, 1]$. Cohen et al.~\cite{Cohen0NSS24} gave a matching lower bound for this problem. These bounds are incomparable with our results since the class of queries (threshold queries) is known ahead of time, and the universe size is infinite.

The privacy analysis in our upper bound is based on a line of work going back to the AboveThreshold/SVT mechanism~\cite{DworkR14}. Later, \cite{KaplanMS21} introduced SVT with individual privacy charging, which can be generalized by the target charging framework of \cite{CohenL23}. 

\section{Preliminaries}\label{sect:prelims}
\subsection{Notational Conventions}
\begin{itemize}
    \item Given $a \in \N$, we use $[a] \coloneq  \set{1,\dots,a}$ to denote the set of the first $a$ natural numbers.
    \item Throughout the paper, $n$ refers to the size of the dataset, $m$ refers to the number of queries and $N$ refers to the size of the universe $\mathcal U$. 
    \item We use $\mathcal{P}_V$ to denote the projection matrix onto a subspace $V$. 
    \item We use $X \succeq 0$ to denote that $X$ is a positive semidefinite (PSD) matrix and $X \succeq Y$ to denote that $X - Y \succeq 0$. The notation $X \preceq Y$ is equivalent to $Y \succeq X$. 
    \item For two quantities $A$ and $B$ that may depend on other parameters, we use the notation $A\lesssim B$ when there is an absolute constant $C$ such that $A \le CB$; the notation $A \gtrsim B$ is equivalent to $B \lesssim A$. 
    \item For a vector $v \in \R^d$, we use the standard notation for the $\ell_1$, $\ell_2$ and $\ell_\infty$ norms:  $$\| v \|_1 \coloneq  {|v_1| + \dots + |v_d|}\qquad\| v \|_2 \coloneq  \sqrt{|v_1|^2 + \dots + |v_d|^2}\qquad\| v \|_{\infty} \coloneq  \max_{i \in [d]} |v_{i}|.$$ 
\end{itemize}

\subsection{Matrix Factorization Norms}

Factorization norms were introduced in functional analysis to study linear operators that factor through Hilbert space. Here, we focus on a special case that is most common in applications in computer science, and give the definitions in terms of matrices. We define the $\gamma_2$ factorization norm of an $m\times N$ matrix $A$ as
\[
\gamma_2(A) = \min\{\|L\|_{2\to\infty} \|R\|_{1\to 2}: LR = A\}.
\]
Recall that $\|R\|_{1\to 2}$ is the maximum $\ell_2$ norm of a column of $R$, and $\|L\|_{2\to\infty}$ is the maximum $\ell_2$ norm of a row of $L$. 


\subsection{Basic Tools for Differential Privacy}
In this subsection we define some of the most well-known tools for differential privacy. 
\begin{lemma}[Laplace mechanism {\cite{DworkR14}}]
Let $f:\mathcal{U}^n\to\mathbb{R}$ have sensitivity at most $\Delta>0$, where 
\[\Delta \coloneq  \sup_{X,X' \text{ neighboring}}\abs{f(X)-f(X')}.\]
For any $\varepsilon>0$, the mechanism
\[
    \mathcal{M}(X)=f(X)+Z,
    \qquad Z\sim\operatorname{Lap}(\Delta/\varepsilon),
\]
is $\varepsilon$-differentially private.
\end{lemma}


\begin{lemma}[Post-processing {\cite{DworkR14}}]
If $\mathcal M$ is $(\varepsilon,\delta)$-differentially private
and $f$ is a mapping that does not depend on $X$,
then $f\circ\mathcal M$ is $(\varepsilon,\delta)$-differentially private.
\end{lemma}

\subsection{Interactive Differential Privacy}

We need to define privacy and accuracy for algorithms interacting with a data analyst. We do so in terms of a meta algorithm that outputs the transcript of the interaction between the algorithm $\mathcal{M}$ and an \emph{interactive data analyst} $A$.

We first consider an oblivious analyst $A$ that determines the sequence of queries given to $\mathcal{M}$ independently of the answers $\mathcal{M}$ gives. Without loss of generality, we can assume that $A$ decides the sequence of queries before the interaction starts, as shown in Algorithm~\ref{fig:transcript-oblivious}. It is worth noting that an oblivious adversary can still know what the mechanism $\mathcal{M}$ is and choose the queries with this knowledge.

\begin{algorithm}[htbp]
\caption{The meta-algorithm
$\mathsf{Trans}_{m}^{\mathcal{M} \leftrightarrow A}(X)$
for oblivious analysts}
\label{fig:transcript-oblivious}
\begin{algorithmic}[1]
\Require Dataset $X \in \mathcal{U}^n$
\State $A$ chooses queries $q_1,\ldots,q_m$.
\For{$j = 1,\ldots,m$}
    \State Give $q_j$ to $\mathcal{M}(X)$ and receive an answer $o_j$.
\EndFor
\State \Return $(Q=(q_1,\ldots,q_m),
                 o=(o_1,\ldots,o_m))$.
\end{algorithmic}
\end{algorithm}

We also consider adaptive analysts $A$ that determine the next query to be given to $\mathcal{M}$ based on the answers to previous queries, as shown in Algorithm~\ref{fig:transcript-adaptive}.

\begin{algorithm}[htbp]
\caption{The meta-algorithm
$\mathsf{Trans}_{m}^{\mathcal{M} \leftrightarrow A}(X)$
for adaptive analysts}
\label{fig:transcript-adaptive}
\begin{algorithmic}[1]
\Require Dataset $X \in \mathcal{U}^n$
\State $o_0 \gets \bot$.
\For{$j = 1,\ldots,m$}
    \State Give $o_{j-1}$ to $A$ and receive a query $q_j$.
    \State Give $q_j$ to $\mathcal{M}(X)$ and receive an answer $o_j$.
\EndFor
\State \Return $(Q=(q_1,\ldots,q_m),
                 o=(o_1,\ldots,o_m))$.
\end{algorithmic}
\end{algorithm}

\begin{defn}[Privacy of Interactive Algorithms]
An interactive algorithm $\mathcal{M}$ is \emph{$(\eps, \delta)$-differentially private} (for non-adaptive/adaptive analysts) if for every dataset $X \in U^n$, and every (resp.~non-adaptive/adaptive) analyst $A$, the algorithm $\mathsf{Trans}_{m}^{\mathcal{M} \prot A}$ is $(\eps, \delta)$-differentially private.
\end{defn}

Similarly, we can define accuracy for interactive algorithms via the same meta-algorithm:

\begin{defn}[($\alpha, \beta$)-Accuracy of Interactive Algorithms]
An interactive algorithm $M$ is \emph{$(\alpha(Q),\beta)$-accurate for $m$ queries} against non-adaptive/adaptive analysts if for every dataset $X \in U^n$, and every (resp.~non-adaptive/adaptive) analyst $A$, if $Q$ is the queries submitted by $A$,
$$
\pr{(Q,o) \gets \mathsf{Trans}_{m}^{M \prot A}(X)}{ \| Q(X) - o \|_{\infty} \leq \alpha(Q)} \geq 1-\beta,
$$
where $\alpha(Q)$ is some real-valued function of $Q$.
\end{defn}

\begin{definition}[$\ell_\infty^2$ Error of Interactive Algorithms]
An interactive algorithm $\mathcal{M}$ has $\ell_\infty^2$ error
at most $\alpha(Q)$ for $m$ queries against oblivious/adaptive analysts
if, for every dataset $X\in\mathcal{U}^n$ and every
(resp.~oblivious/adaptive) analyst $A$,
\[
\left(
    \mathbb{E}_{(Q,o)\gets
        \mathsf{Trans}_{m}^{\mathcal{M}\prot A}(X)}
    \left[
        \|Q(X)-o\|_\infty^2
    \right]
\right)^{1/2}
\leq \alpha(Q),
\]
where $\alpha(Q)$ is some real-valued function of $Q$. The expectation is over the randomness of both $\mathcal{M}$ and $A$.
Moreover, we define the $\ell_\infty^2$ error for the worst case dataset to be
\[
\mathrm{err}(\mathcal{M},A)\coloneq 
\sup_{X}
\left(
    \mathbb{E}_{(Q,o)\gets
        \mathsf{Trans}_{m}^{\mathcal{M}\prot A}(X)}
    \left[
        \|Q(X)-o\|_\infty^2
    \right]
\right)^{1/2}.
\]
Moreover, when the adversary is oblivious, since the queries do not depend on the answers, we can denote the error as $\mathrm{err}(\mathcal{M},Q)$.
\end{definition}

We remark that the supremum in $\ell_\infty^2$ error is taken over all datasets of \textit{arbitrary size}.

\section{An Equivalent Form for Online Factorization Algorithms}
In this technical section, we define the Growing Matrix Problem, which will be a convenient alternative way to view the online factorization problem. 
\begin{definition}[The Growing Matrix Problem]
    For a query matrix $Q$ arriving online, we say an online algorithm that outputs a sequence of positive semidefinite (PSD) matrices $\Sigma_1, \ldots, \Sigma_m$ achieves value $\gamma^2$ for the Growing Matrix Problem over $Q$ if, for each $t \in [m]$,
    \begin{itemize}
        \item $\Sigma_t$ is a $t \times t$ PSD matrix;
        \item $\Sigma_t$ is the principal submatrix of $\Sigma_{t+1}$ consisting of its top $t$ rows and columns;
        \item $\Sigma_m$ (and, therefore, every $\Sigma_t$) has all its diagonal entries bounded by $\gamma^2$, for a non-negative number $\gamma$;
        \item if $Q_t$ is the submatrix of $Q$ that has arrived by time step $t$, and $q$ is a column of $Q_t$, then $\Sigma_t \succeq q q^T$.
    \end{itemize}
    
\end{definition}
The Proposition below states the equivalence between the Growing Matrix Problem and the online factorization problem. At a high level, $\Sigma_t$ corresponds to $L_tL_t^T$, and its growing property allows us to construct the $R_t$ matrix so that only new rows are being appended to it over time.
\newcommand{\growingMatrixStatement}{%
    There is an online algorithm that achieves value $\gamma^2$ for the Growing Matrix Problem over $Q$ if and only if there exists an online factorization mechanism for $Q$ with factorization value $\gamma$.
}
\begin{proposition}\label{prop:SDP}
\growingMatrixStatement
\end{proposition}
The proof of the above proposition can be found in \hyperlink{app:prop:SDP}{Appendix~\ref*{app:proofs}}. Let us also prove the following corollary of Proposition~\ref{prop:SDP}, which will be used later in Section~\ref{sect:upper_bound}.
\begin{corollary}\label{lem:one_row_per_time}
    For any online factorization algorithm, there exists an equivalent online factorization algorithm that, at each time $t$, appends at most one row to $R_{t-1}$.
\end{corollary}
\begin{proof}
    Take any online factorization algorithm, and use Proposition~\ref{prop:SDP} to first get an online algorithm for the growing matrix problem, and then use the proposition again to get back an equivalent online factorization. By construction, we increase the number of rows of $M_t$ by only one at each time step $t$, and therefore, we also increase the number of rows of $R_t = M_t Q_t$ by at most $1$.
\end{proof}

\section{Lower Bound on Error for Private Online Query Release}\label{sect:lower_bound}

In this section, we show a lower bound similar to \cite[Theorem~30]{EdmondsNU20}, but for online private query release algorithms. We will show that any private interactive mechanism can be made data-independent when the size of the dataset is large enough, with a slightly modified version of a result in \cite{BDKT12}, and that any data-independent private interactive mechanism implies an online factorization algorithm by using the data-independent noise's covariance as the growing matrix, which implies an online factorization. Then we prove a lower bound on the competitive ratio of online factorization by looking at the family of threshold queries corresponding to binary search, and show that at least one search path has to have a lower bounded competitive ratio.

We start by showing the online version of Theorem 30 in~\cite{EdmondsNU20}.

\begin{lemma}\label{lem:PrivAlg_to_RandFac}
Let $\mathcal{M}$ be an online $(\eps,\delta)$-differentially private algorithm answering obliviously chosen linear queries on datasets $X\in \mathcal{U}^n$. For every $\eps, \delta$ smaller than some absolute constants, there exists a deterministic online factorization algorithm that on every query matrix $Q$ achieves factorization value
\[
    \gamma=O\left(
\eps\cdot
        \mathrm{err}(\mathcal{M},Q)
        \right).
\]
\end{lemma}
As in the original proof, we first need to apply a reduction from data-dependent algorithms to data independent ones. This is \cite[Lemma~29]{EdmondsNU20}, a slightly modified version of \cite[Theorem~4.3]{BDKT12}, which also holds against oblivious online queries. Let us restate it here.

\newcommand{\bdktStatement}{%
    For any oblivious sequence of linear queries with query matrix $Q\in \R^{m\times N}$, for every interactive $(\eps,\delta)$-differentially private mechanism $\mathcal{M}$, and for any $c>0$, there exists an $(\eps, \delta+c)$-differentially private data-independent mechanism $\mathcal{M'}$ such that 
    \[
        \mathrm{err}(\mathcal{M'},Q)
        \leq
        \mathrm{err}(\mathcal{M},Q).
    \] 
    Here, we say an online algorithm is data-independent if, upon receiving query $q_t$ at time $t$, it outputs $q_t(X)+\mathcal{W}$ where $\mathcal{W}$ does not depend on $X$.
}
\begin{lemma}[Online version of Lemma~29, \cite{EdmondsNU20}]\label{lem:bdkt}
\bdktStatement
\end{lemma}
The proof of the above lemma can be found in \hyperlink{app:lem:bdkt}{Appendix~\ref*{app:proofs}} - it follows the same idea as the proof of \cite[Theorem~4.3]{BDKT12}, but uses a different noise distribution and a simpler analysis. Now let us return to Lemma~\ref{lem:PrivAlg_to_RandFac}.
\begin{proof}[Proof of Lemma~\ref{lem:PrivAlg_to_RandFac}]
    First we note that the original theorem was for $\ell_2$ error, but as mentioned right below the original proof it also extends to the max error. Applying Lemma~\ref{lem:bdkt} we obtain an online noise adding algorithm, at each time $t$ adding noise $Z_t$, independent of the dataset $X$, to $q_t(X)$. For some absolute constant $C$, we take
    \[
    \Sigma_t = C^2\eps^2\operatorname{Cov}(Z_{\leq t}),
    \]
    where $Z_{\leq t}$ contains all the noises up to time $t$. By the same covariance bound in \cite[Lemma~27]{EdmondsNU20}, we obtain 
    \[
    \Sigma_t\succeq qq^{\mathsf T}.
    \]
    Furthermore, we can bound the diagonal entries
    \[
    (\Sigma_t)_{ii}=C^2\eps^2\operatorname{Var}(Z_i)\lesssim \left(\eps\cdot
        \mathrm{err}(\mathcal{M},Q)\right)^2
    \]
    Applying Proposition~\ref{prop:SDP} finishes the proof.
\end{proof}

We now have a reduction from general online private query release to deterministic online factorization. We will give a direct lower bound for the latter. We consider a binary search tree, and, for any path in the tree, the sequence of threshold queries associated with nodes on the path, where threshold queries are a special subset of linear queries, defined as follows:
\begin{definition}[Threshold Query]
    A threshold query is specified by
a function $q^\nu:\mathcal{U}\to\{0,1\}$ such that \[q^\nu(x)=\mathbf{1}\{x\le \nu\}.\] For a dataset
$X=(x_1,\ldots,x_n)\in\mathcal{U}^n$, its answer is
\[
    q^\nu(X) \coloneq  \sum_{i=1}^{n} q^\nu(x_i).
\]
\end{definition}
We show that no factorization algorithm can achieve a very small factorization value on all such paths. The key idea of the proof is that, at each step of choosing whether to ask a query corresponding to the left or right side of the current search range, there must be one side that makes the average column norm of $R$ restricted to the search range large, and thus one can always choose that side as the next query. 

\begin{lemma}\label{lem:lb_base}
    For threshold queries on domain $[N]$: assume $N$ is a power of 2 and construct a full binary tree, where the root node represents $[N]$, its children represent $[N/2]$ and $[N]\setminus[N/2]$, etc., down to the leaf nodes that represent singleton sets $\{i\}$ for $i \in [N]$. We let each node represent the threshold query indexed by the middle point of the range it represents. Then each path from root to leaf forms a sequence of queries. Then, for any deterministic online factorization algorithm, there exists a path on which the algorithm achieves factorization value $\Omega(\sqrt{\log{N}})$.

    As a direct consequence, the worst case competitive ratio of any deterministic online factorization algorithm  is $\Omega(\frac{\sqrt{\log N}}{\log\log N})$.
\end{lemma}
\begin{proof}
Since we are considering deterministic factorization algorithms, it is well-known (\cite{Karp92}) that adaptively chosen inputs have no extra power compared with oblivious inputs. So, we may assume without loss of generality that the queries are chosen adversarially. Consider the following adaptive adversary: At time $t$, after asking the threshold query corresponding to a range $S$, the adversary selects one of $S_L$ and $S_R$, the left and right disjoint halves of $S$. Let us define 
\[
\mu_t(S)\coloneq \frac{R_t\mathbf{1}_S}{|S|}.
\]
The above vector is an average of the columns of $R_t$ inside the range $S$. 
We will show that $\max\{\|\mu_t(S_L)\|_2, \|\mu_t(S_R)\|_2\}$ is strictly larger than $\|\mu_t(S)\|_2$. We know that $\ell_tR_t=q_t$, where $q_t$ is the threshold query corresponding to the range $S$. By the definition of a threshold query, $q_t$ is a binary vector that has all ones up to the middle point of $S$, and all zeros after that. Then it is clear that
\[
\ell_t\mu_t(S_L)=1,\qquad \ell_t\mu_t(S_R)=0,
\]
and, therefore, by Cauchy-Schwarz, 
\[
\norm{\mu_t(S_L)-\mu_t(S_R)}_2 \ge \frac{\ell_t(\mu_t(S_L)-\mu_t(S_R))}{\norm{\ell_t}_2} \ge \frac{1}{\norm{\ell_t}_2}.
\]
By the parallelogram identity, we have 
\begin{align*}
\frac{\norm{\mu_t(S_L)}_2^2+\norm{\mu_t(S_R)}_2^2}{2}
    &=\norm{\mu_t(S)}_2^2+\frac{1}{4}\norm{\mu_t(S_L)-\mu_t(S_R)}_2^2\\
    &\geq \norm{\mu_t(S)}_2^2+\frac{1}{4\norm{\ell_t}_2^2}
\end{align*}
Thus at each time $t$ there exists $b\in \{L, R\}$ s.t. \[\norm{\mu_t(S_b)}_2^2\geq \norm{\mu_t(S)}_2^2+\frac{1}{4\norm{\ell_t}_2^2}.\]
At each step, the adversary chooses this $b$, and the average column norm squared over the current range grows by an additive factor of at least $\frac{1}{4\norm{\ell_t}_2^2}$. Note that the online factorization algorithm only appends rows to the $R_t$ matrix. Thus when the leaf node is reached, since each leaf node corresponds to only one element, the column norm corresponding to this leaf node with range $S_{final}$ is
\[
\norm{\mu_t(S_{final})}_2^2\geq \frac{\log{N}}{4\norm{L}_{2\rightarrow \infty}^2}.
\]
Hence, denoting the number of queries by $m$, we have \[
\norm{R_m}_{1\rightarrow 2}\norm{L}_{2\rightarrow \infty}\geq \frac{\sqrt{\log{N}}}{2}.
\]
To prove the second part of the lemma, observe that any path of the tree corresponds to a sequence of $m = O(\log N)$ threshold queries, and the $\gamma_2$ norm of the corresponding query matrix is known to be only $O(\log m) = O(\log \log N)$ by the work \cite{MatousekNT14}.
\end{proof}
We are now fully equipped with all the tools we need: Combining Lemma~\ref{lem:lb_base} and Lemma~\ref{lem:PrivAlg_to_RandFac} proves our lower bound theorem.

\begin{lowerboundmain}[Main Lower Bound, Formal]\label{thm:main_LB_full}
    For sufficiently small privacy parameters $\varepsilon,\delta$
bounded by absolute constants, for any $(\eps,\delta)$-DP online mechanism $\mathcal{M}$, there exists a sequence of $m$ obliviously chosen linear queries $Q=(q_1,...,q_m)$ over a universe of size $N$, such that 
\[
    \mathrm{err}(\mathcal{M},Q)=\Omega\left(\frac{\gamma_2(Q)\sqrt{\log N}}{C\eps\log\log N}
        \right).
\]
\begin{rem}
    One can show that for sufficiently small $\eps,\delta,\beta$, no $(\eps,\delta+\beta)$-differentially private algorithm can achieve error 
    \[
        O\!\left(\frac{\gamma_2(Q)\sqrt{\log N}}{C\eps\log\log N}
        \right),
    \] with probability more than $1-\beta$ on all sequences of queries, by clipping the noise $Z$ after applying Lemma~\ref{lem:bdkt}.
\end{rem}
\end{lowerboundmain}
\section{Adaptive Private Query Release}\label{sect:upper_bound}

In this section, we consider the following problem: we hold a private point $x\in \R^d$, where neighboring points have $\ell_2$ distance at most 1 from each other. At each time step $t$, we receive a query $q_t$. We want to release a private estimate of $\iprod{q_t}{x}$, and we give accuracy bounds assuming queries have $\ell_2$ norm at most $\gamma$. In this section, we design a layered Iterative Database Construction algorithm, that keeps a public synthetic point to answer the queries, and updates it according to the queries. The algorithm is composed of several Euclidean IDC instances, each handling a subspace of the data universe, where the subspaces for different instances are pairwise orthogonal. Before any query arrives, the first layer handles the entire space, while other layers handle the empty subspace. We use a kick-out trick to maintain bounded privacy and error for each instance - whenever the public dataset is updated, the span of the direction of update, which is the direction of the query vector projected to the subspace handled by the current layer, will be moved to the next layer. Apart from that, within each layer, the private test and the subsequent update can be viewed as a private mechanism together. We call this bundled mechanism $\operatorname{InsideOrRelease}$. Then the execution of our algorithm will make a sequence of calls to $\operatorname{InsideOrRelease}$, which can be analyzed using the target charging framework. One nuance is that these individual calls will have non-uniform privacy parameters that depend on the previous calls, so we need to provide an extended version of the target charging analysis suited to our use case.

\subsection{The Target Charging Framework}
Here we introduce some elements of the target charging analysis by Cohen and Lyu \cite{CohenL23}, which our privacy analysis will rely on.
\begin{definition}[$q$-target {\cite[Definition~2.1]{CohenL23}}]
\label{def:q_target}
Let $\mathcal{A}\colon\mathcal{X}\to\mathcal{Y}$ be a randomized
algorithm, and let $q\in(0,1], \varepsilon\ge0$.
A measurable set $\mathcal{T}\subseteq\mathcal{Y}$ is a
\emph{$q$-target with parameter $\eps$} for $\mathcal{A}$
if, for every neighboring pair $x_0,x_1\in\mathcal{X}$, there exist
$r\in[0,1]$, a distribution $C$ on $\mathcal{Y}$, and an
$\eps$-DP mechanism
$\mathcal{M}\colon\{0,1\}\to\mathcal{Y}$, such that if $P_j, Q_j$ are the distributions of $\mathcal{A}(x_j), \mathcal{M}(j)$ respectively, then for every $j \in \{0,1\}$, 
\[
    P_j
    =
    (1-r)C+r\,Q_j,
    \qquad
    \Pr[\mathcal{M}(j)\in\mathcal{T}]\ge q.
\]
An output in $\mathcal{T}$ is called a \emph{target hit}.
\end{definition}
The target charging framework gives a convenient way for proving privacy of a composition of a collection of private algorithms when there is a guaranteed bound on the number of target hits. Each algorithm can be viewed as consisting of two branches, one branch that does not use the dataset at all, and another branch that depends on the dataset. A target is a subset $\mathcal{T}$ of the output space such that when the algorithm goes into the branch that uses the dataset, it is guaranteed to produce an output in  $\mathcal{T}$ with probability bounded from below. Intuitively, if we have an upper bound on the number of target hits, we can bound the number of times the algorithms go into the data-dependent branch and thus also bound the overall privacy loss of the composition. This clearly gives a tighter analysis than simply applying a composition theorem to the entire collection of algorithms, and ignoring the bound on the number of target hits. For this paper, we only use one kind of target proposed in \cite{CohenL23}, restated in the following lemma.
\begin{lemma}[NotPrior target
    {\cite[Lemma~2.3; Lemma~C.2]{CohenL23}}]
\label{lem:notprior_target}
Let $\mathcal{A}\colon\mathcal{X}\to
\mathcal{Y}\cup\{\bot\}$ be an $\eps$-DP algorithm,
where $\mathcal{Y}$ is nonempty and $\bot\notin\mathcal{Y}$.
Then $\mathcal{Y}$ is a
    ${1/(e^{\eps}+1})$-target with parameter $\eps$ for $\mathcal{A}$.

The lemma holds when the input domain is restricted to a fixed neighboring pair. 
\end{lemma}


\subsection{Target Charging Analysis for Private Threshold Tests}
Let us first define the following subroutine that is being used in our algorithm. For $b>0$, a unit direction $u$, and public thresholds $\tau_-<\tau_+$, we define $\mathsf{InsideOrRelease}_b(x;u,\tau_-,\tau_+)$ to be a noisy test for whether $\iprod{x}{u}$ lands inside $[\tau_-,\tau_+]$. When this noisy test does not report $\mathsf{Inside}$, it outputs a noisy estimate of $\iprod{x}{u}$. The parameter $b$ controls the magnitude of noises. Every call to this algorithm uses fresh independent randomness.
\begin{algorithm}[H]
\caption{$\mathsf{InsideOrRelease}_b(x;u,\tau_-,\tau_+)$}
\label{alg:inside_or_release}
\begin{algorithmic}[1]
    \Statex \textbf{Input:}
        $x\in\mathbb{R}^d$, $\|u\|_2=1$,
        $\tau_-<\tau_+$, $b>0$.
    \State Draw $Z\sim\operatorname{Lap}(0,b)$
    \If{$\tau_-<\iprod{x}{u}+Z<\tau_+$}
        \State \Return $\mathsf{Inside}$.
    \Else
        \State \Return $\langle x,u\rangle+Z$.
    \EndIf
\end{algorithmic}
\end{algorithm}
We now state the privacy guarantee of this subroutine.

\begin{lemma}[Indistinguishability of $\mathsf{InsideOrRelease}$]\label{lem:inside_or_release}
    Fix $u, \tau_-, \tau_+$ and a pair $x, x'$. The outputs of Algorithm~\ref{alg:inside_or_release} over $x$ and $x'$ are $\frac{\abs{\iprod{x-x'}{u}}}{b}$-indistinguishable, where we say two distributions $P$ and $P'$ are $\eps$-indistinguishable if for any measurable subset $S$,
    \[e^{-\eps} P'(S)\leq P(S)\leq e^\eps P'(S).\]
\end{lemma}
\begin{proof}
    This lemma follows trivially from privacy guarantees of the Laplace noise mechanisms and postprocessing. 
\end{proof}
This lemma is equivalent to saying the subroutine is differentially private up to the same parameter for fixed neighboring pairs, which is what Lemma~\ref{lem:notprior_target} uses. The target will be $\R$, i.e. when the above test releases a real number. In fact, later in this section we choose appropriate parameters so that $\R$ is a $1/4$-target.

For the purpose of our algorithm, which is to be defined in a later subsection, we need to analyze the composition of a collection of private algorithms with non-uniform privacy parameters, and these parameters could depend on the execution of the previous algorithms.
This is an extended version of \cite[Lemma~2.2; Remark~B.6]{CohenL23}.

\begin{algorithm}[H]
\caption{Weighted Target Charging with bounded privacy loss}
\label{alg:weighted_target_charging}
\begin{algorithmic}[1]
    \Statex \textbf{Input:}
        Dataset $x\in\mathcal{X}$;
        horizon $T\in\mathbb{N}$.

    \For{$i=1,\ldots,T$}
        \State Receive $(\mathcal{A}_i,\top_i)$,
            chosen according to the preceding transcript,
        \Statex \hspace{\algorithmicindent}
            where $\mathcal{A}_i$ is $\eps_i$-DP, 
            $\eps_i$ is determined by the history before time $i$,
            
        \Statex \hspace{\algorithmicindent}
            and $\top_i$ is a $q$-target with parameter
            $\eps_i$ for $\mathcal{A}_i$.
        \State Release $\mathcal{A}_i(x)$
    \EndFor
\end{algorithmic}
\end{algorithm}
\newcommand{\weightedTargetChargingStatement}{%
For every $\eta\in(0,1)$ and $\alpha>0$, if in every execution,
\[
\sum_{i\text{ s.t. }\mathcal{A}_i(x)\in \top_i}\eps_i^2\leq S, \qquad q\in(0,1], \qquad   0\leq \eps_i\leq\bar\eps\leq 1,
\]
then Algorithm~\ref{alg:weighted_target_charging} is
\[
    \left(
        \frac{(1+\alpha)S}{2q}
        +
        \sqrt{2\log(1/\eta)(1+\alpha)\frac{S}{q}},
        \;
        \eta+\exp\left(
        \left(
        \log{(1+\alpha)}-\alpha
        \right)
        \frac{S}{\bar\eps^2}
    \right)
    \right)\text{-DP}.
\]
}
\begin{lemma}[Privacy of weighted target charging with bounded privacy loss]
\label{lem:weighted_target_charging}
\weightedTargetChargingStatement
\end{lemma}
We defer the proof to \hyperlink{app:lem:weighted_target_charging}{Appendix~\ref*{app:proofs}}.

\subsection{Iterative Database Construction with Layered Subspaces}
Now we define the layered IDC algorithm, as shown in Algorithm~\ref{alg:layered_idc}. The algorithm holds a public synthetic point $\hat{x}$, where initially $\hat{x} = 0$. We partition the data universe into orthogonal subspaces $V_1,V_2,...,$. Initially, $V_1=\R^d$ and $\forall j>1, V_j=\{0\}$. On each query $q_t$, the algorithm goes through every $V_j$, and tries to examine if $\hat{x}$ and $x$ are far in the direction of $u \coloneq  {\mathcal{P}_{V_j}q_t}/{\norm{\mathcal{P}_{V_j}q_t}_2}$, i.e. $q_t$ projected to subspace $V_j$ and renormalized. If they are close, $\hat{x}$ can be used to answer the query inside this subspace. Otherwise, we release a noisy estimate $z=\iprod{x}{u}+Z$. We shift $\hat{x}$ alongside $u$ such that $\iprod{\hat{x}}{u}=z$, and kick out the line through direction $u$ to the next subspace. Effectively, after identifying a bad direction, we pay some privacy cost to rewrite $\hat{x}$ only in this direction. Each of these kick-out events can be viewed as a target hit, and we will use the target charging technique for the analysis. 

The layered structure gives several desirable properties. By separating the subspaces orthogonally, future repairs in the remaining directions will not damage what is already repaired in the current layer. We also bound the number of layers needed by a potential argument. For each layer $j$, consider the potential \[
\Phi_j=\|\mathcal{P}_{V_j}(x-\hat{x})\|_2^2,
\]  
which is the distance between the public and private points inside subspace $V_j$. When we kick out a direction, we can give an upper bound on the potential increment in the next layer (i.e., the layer receiving the kicked out direction) because this distance is exactly the square of the noise inside the estimate we just released. We can also give a lower bound on the potential decrement in the current layer because, to trigger an update, $x$ and $\hat{x}$ must have been sufficiently separated in the query direction, and the contribution of this direction to the potential will become 0, since it no longer belongs to the current layer. We can then show the total potential gain for each layer decreases geometrically from one layer to the next, which allows us to bound the number of layers.

\begin{algorithm}[htp]
\caption{Layered IDC}
\label{alg:layered_idc}
\begin{algorithmic}[1]
    \Require Private point $x\in\mathbb{R}^d$; Query horizon $m$;
        parameters $a,b>0$.
    \State $\hat{x}\gets0\in\mathbb{R}^d$
        \Comment{Public synthetic point}
    \State $V_1\gets\mathbb{R}^d$
    \State $V_2,\ldots,V_{L+1}\gets\{0\}$
        \Comment{The final layer $V_{L+1}$ is never tested}

    \For{$t=1,\ldots,m$}
        \State Receive $q_t\in\mathbb{R}^d$

        \For{$j=1,\ldots,L$}
            \State $v\gets\mathcal{P}_{V_j}q_t$
                \Comment{Projection onto the current layer}
            \If{$\|v\|_2>0$}
                \State $u\gets v/\|v\|_2$;
                \State $z\gets
                    \mathsf{InsideOrRelease}_b
                    (x;u,\iprod{\hat{x}}{u}-2a,\iprod{\hat{x}}{u}+2a)$
                \If{$z\neq\mathsf{Inside}$}
                    \State $\hat{x}\gets \hat{x}+(z-\iprod{\hat{x}}{u})u$
                    \State $V_j\gets V_j\cap u^\perp$
                        \Comment{Kick $u$ out of layer $j$}
                    \State $V_{j+1}\gets
                        V_{j+1}\oplus\operatorname{span}\{u\}$
                \EndIf
            \EndIf
        \EndFor

        \State Release $o_t\coloneq \langle \hat{x},q_t\rangle$
    \EndFor
\end{algorithmic}
\end{algorithm}
The parameters $a$ and $b$ are for controlling, respectively, the width of the inside range and the magnitude of the noise of~$\mathsf{InsideOrRelease}$. Let us first prove the following technical lemma which bounds the privacy cost of all the target hits. Recall that a target hit here is just when $\mathsf{InsideOrRelease}$ outputs a real number, which triggers an update to the public point.
\begin{lemma}[Orthonormal Properties and Counting kick-outs]
\label{lem:counting_kickouts}
In Algorithm~\ref{alg:layered_idc}, index all calls to
$\mathsf{InsideOrRelease}_b$ by $i$, and let $u_i$ be the
unit direction used in call $i$. For each layer $j\leq L$,
let $I_j$ be the set of indices of calls that kick a direction
out of $V_j$. On every execution, the directions
$\{u_i:i\in I_j\}$ are orthonormal within each layer
$j\in\{1,\ldots,L\}$.

Fix neighboring points $x,x'\in\mathbb{R}^d$, and assign call $i$ the pairwise privacy parameter
\[
    \eps_i
    =
    \frac{|\langle x-x',u_i\rangle|}{b}.
\]
Then $0\le\eps_i\le1/b$, and on every execution,
\[
    \sum_{j=1}^{L}\sum_{i\in I_j}\eps_i^2
    \le
    \frac{L}{b^2}.
\]
Equivalently, since each numerical output is a target hit and
causes exactly one kick-out,
\[
    \sum_{i:\,\text{target hit}}\eps_i^2
    \le
    \frac{L}{b^2}.
\]
\end{lemma}
\begin{proof}
    The orthonormal property holds because, once $u_i$ is kicked out of $V_j$, all future directions $u_{k}$ used to call $\mathsf{InsideOrRelease}_b$ inside layer $j$ lie inside $V_j \cap u_i^\perp$. The fact that $\eps_i\leq 1/b$ is trivial from the parameter choice. To prove the rest of this lemma, observe that the orthonormal property implies that each layer may only contribute at most $\frac{1}{b^2}\|x-x'\|_2^2$ to the sum.
\end{proof}

We can now prove the main upper bound theorem for the layered IDC.
Lemma~\ref{lem:counting_kickouts} gives an upper bound on the privacy cost of target hits, i.e., the iterations in which $\mathsf{InsideOrRelease}_b$ outputs a number. With the target charging analysis, this will allow us to bound the overall privacy loss of the algorithm. The layered structure will ensure the error contributed by each layer is bounded. We will show that no element ever enters the last, never-tested layer, via a potential argument, thus bounding the overall error.
\begin{theorem}[Privacy and accuracy of Layered IDC]
\label{thm:layered_idc}
Let $m\ge1$, $\gamma>0$, $\eps>0$, and
$\delta>0,\beta\in (0, 1)$.
For a sufficiently large universal constant $C>0$, set
\[
    L=2+\lfloor\log_9 d\rfloor,
    \qquad
    b=\max\left(1, C\sqrt{\frac{L}{\eps}},\frac{C}{\eps}\left(\sqrt{L\log\frac{1}{\delta}}+\log\frac{1}{\delta}\right)\right),
    \qquad
    a=2b\log\frac{mL}{\beta}.
\]

Then Algorithm~\ref{alg:layered_idc} is
$(\eps,\delta)$-differentially private. Moreover, for every $x\in\mathbb{R}^d$ and
every adaptive analyst submitting at most $m$ queries
$q_t\in\mathbb{R}^d$ with $\|q_t\|_2\le\gamma$,
\[
    \Pr\left[
        \max_{t\le m}
        |o_t-\langle x,q_t\rangle|
        \le
        \frac{5\gamma a\sqrt{L}}{2}
    \right]
    \ge 1-\beta.
\]
\end{theorem}
\begin{proof}
    Since, by Lemma~\ref{lem:notprior_target}, $\eps_i\leq 1/b\leq 1$, each real output of $\mathsf{InsideOrRelease}$ is a $1/4$-target. Combining Lemma~\ref{lem:counting_kickouts} with Lemma~\ref{lem:weighted_target_charging}, and choosing $$S=\frac{L}{b^2},\qquad \alpha=\frac{4\log{\frac{1}{\delta}}}{L}+2,\qquad \eta=\delta/2$$ proves the privacy guarantee. To understand the intuition behind the parameter choices here and in the theorem statement, the choice of $\alpha$ bounds the failure probability, i.e. the $\delta$ term; the first term of $b$ is to bound each $\eps_i$ by $1$; the other two terms are to give us the $\eps$ term in $(\eps,\delta)$-differential privacy.

    Now we prove the error bound. There are at most $mL$ Laplace draws, and each satisfies $\Pr(|Z|\ge a/2)=e^{-a/(2b)}$. Thus, by the choice of $a$ and a union bound, all the Laplace noises have absolute value less than $a/2$ with probability at least $1-\beta$. We bound the error assuming this event happens. 
    
    For each layer $j$, recall that the potential is \[
\Phi_j=\|\mathcal{P}_{V_j}(x-\hat{x})\|_2^2.
\]
Observe that whenever the kick-out happens at layer $j$, $\Phi_j$ drops by at least $\frac{9a^2}{4}$, and $\Phi_{j+1}$ increases by at most $\frac{a^2}{4}$. Since $\Phi_j$ is non-negative at every time step, this means the number of dimensions that enter each layer $j$, $j \ge 3$, is at most $1/9$-th of the number of dimensions that enter the previous layer $j-1$. Moreover, at most $d$ dimensions can enter layer $2$: Layer 1 has unbounded potential but only $d$ dimensions to be kicked out of it. Consequently, at most $d/9^{L-1}<1$ directions can enter layer $L+1$, so this layer must always be empty. 

At any fixed layer $j\leq L$, whenever $\mathsf{InsideOrRelease}$ outputs $\mathsf{Inside}$, or immediately \textbf{after} a kick-out is performed, it is clear that \[\abs{\iprod{\hat{x}-x}{u}}\leq \frac{5a}{2}.\] Summing over layers using the orthogonality between layers and then scaling by $\gamma$ finishes the proof.
\end{proof}
\begin{rem}
    In particular, the maximum error is
\[
    O\!\left(
       \gamma
        \left[
        \frac{\log d\sqrt{\log(1/\delta)}
            +
            \sqrt{\log d}\,\log(1/\delta)}{\eps}
            +\frac{\log d}{\sqrt{\eps}}+\sqrt{\log d}
        \right]
        \log\frac{m\log d}{\beta}
    \right)
\]
with probability at least $1-\beta$, and we can take $\beta=\delta$.
\end{rem}
We can easily extend the above theorem to the following setup where new dimensions are added to the private point over time. We will require this variant in our later use of the algorithm.
\begin{cor}\label{cor:layered_idc}
Algorithm~\ref{alg:layered_idc} works if the private point $x$ has growing dimension. In particular, for any time $t$, the algorithm receives a pair $(q_t, x_t)$ such that there exists $U_t\subseteq [d]$, \[U_{t-1}\subseteq U_t,\qquad x_t|_{U_{t-1}}=x_{t-1}|_{U_{t-1}}, \qquad q_t|_{[d]\setminus U_t}=\mathbf{0}\]
The privacy and accuracy guarantees shown above hold exactly unchanged, assuming the neighboring constraints hold for the fully revealed point at the end of the time horizon. 
\end{cor}
\begin{proof}
    We can view the private point $x \coloneq x_m$ as being fixed ahead of time, and the dimensions in $U_t \setminus U_{t-1}$ are introduced into layer $1$ at time step $t$. At most $d$ dimensions are ever introduced into layer $1$. It is easy to verify that the analysis of our mechanism still holds.
\end{proof}
 
\subsection{The Main Algorithm}
In this subsection, we describe how to use the algorithm for Adaptive Inner Product Query Release to answer linear queries without an assumption on the norm of each query. 

Let us recall the adaptive linear query release setup: we hold a private histogram $h\in \R^N$, where neighboring histograms differ at most by 1 at one entry. At each time we receive a query $q_t$ and we want to release a private estimate of $\iprod{q_t}{h}$.

We will use the online factorization algorithm in \cite[Theorem~1.4]{NTU26} and the version of layered IDC in Corollary~\ref{cor:layered_idc} as subroutines. At the arrival of each query, our algorithm uses the online factorization algorithm to maintain a factorization $\ell_tR_t=q_t$, and uses $R_th$ as the point $x$, $\ell_t$ as the query for our IDC algorithm (with the extension introduced in Corollary~\ref{cor:layered_idc}).
\begin{algorithm}[H]
\caption{Online Query Release}
\label{alg:main}
\begin{algorithmic}[1]
    \Require Private histogram $h\in\mathbb{R}^N$; Query horizon $m$; $a,b,L$.
    \State Initialize an instance of $\mathsf{OnlineFactorization}$ with parameter $c$
    \State Initialize an instance of $\mathsf{LayeredIDC}$ with parameters $a, b, L$, using $\R^m$ as the data universe, starting with a point of 0 dimension
    \For{$t=1,\ldots,m$}
        \State Receive $q_t\in\mathbb{R}^N$
        \State Input $q_t$ to $\mathsf{OnlineFactorization}$
        \State $\ell_t,R_t \gets $ output of $\mathsf{OnlineFactorization}$
        \State Input $(\ell_t, R_th)$ to $\mathsf{LayeredIDC}$, receive output ${o_t}$.
        \State Release $o_t$.
    \EndFor
\end{algorithmic}
\end{algorithm}
We now prove the privacy and accuracy of this algorithm.
\begin{upperboundmain}[Main Upper Bound, Formal]\label{thm:main_UB_full}
    Let $m\ge1$, $\gamma>0$, $\eps>0$, and
$\delta>0,c>0,\beta\in (0, 1)$.
For a sufficiently large universal constant $C>0$, set
\[
    L=2+\lfloor\log_9 m\rfloor,
    \qquad
    b=\max\left(1, C\sqrt{\frac{L}{\eps}},\frac{C}{\eps}\left(\sqrt{L\log\frac{1}{\delta}}+\log\frac{1}{\delta}\right)\right),
    \qquad
    a=2b\log\frac{mL}{\beta}.
\]
Algorithm~\ref{alg:main} is
$(\eps,\delta)$-differentially private. Moreover, for every $h\in\mathbb{R}^N$ and
every adaptive analyst submitting at most $m$ queries
$q_t\in\mathbb{R}^N$, define $Q$ to be the query matrix of $m$ rows formed by $q_t$'s. Then, with probability at least $1-\beta$, 
\[
    \max_{t\le m}
        |o_t-\langle h,q_t\rangle|= O\!\left(
            \gamma
        \left[
        \frac{\log m\sqrt{\log(1/\delta)}
            +
            \sqrt{\log m}\,\log(1/\delta)}{\eps}
            +\frac{\log m}{\sqrt{\eps}}+\sqrt{\log m}
        \right]
        \log\frac{m\log m}{\beta}
    \right)
\]
where \[\gamma
    =
    \Theta\!\left(
        \gamma_2(Q)\,
            \log^{\frac12+c}\left(
                \frac{\gamma_2(Q)}{\|q_1\|_\infty}
            \right)
        \log^3 N
    \right).\]
\end{upperboundmain}
\begin{proof}
    By Lemma~\ref{lem:online_factorization_ntu}, $R_th$ is a ``growing'' point for $\mathsf{LayeredIDC}$ and for neighboring $h, h'$
    \[\norm{R_th-R_th'}_2\leq 1, \qquad \norm{\ell_t}_2\leq \gamma.\]
    By Corollary~\ref{lem:one_row_per_time}, we may assume $R_th$ has at most $m$ rows. Thus applying Theorem~\ref{thm:layered_idc} finishes the proof.
 \end{proof}

\subsection{Non-private Adaptive Query Release}
In this subsection we show a direct consequence of the above algorithm for answering statistical queries without privacy. Let us first define statistical queries.
\begin{definition}[Statistical Query]
    A statistical query is specified by
a function $q:\mathcal{U}\to{\{0,1\}}$. For a dataset
$X=(x_1,\ldots,x_n)\in\mathcal{U}^n$, its answer is
\[
    q(X) \coloneq  \frac{1}{n}\sum_{i=1}^{n} q(x_i).
\]
\end{definition}
These statistical queries are simply \textit{normalized} linear queries defined by binary-valued functions. For the matrix representation, the corresponding $Q$ will simply be binary, and the answer will be $Q(X)=\frac{Qh}{n}$. 
In this setting, we assume that each datapoint $x_i$ inside $X$ is drawn from some unknown underlying distribution $\mathcal P$ over $\mathcal U$. We say answers to queries are accurate with respect to the population if we use $\E_{x\sim\mathcal P}[q(x)]$ as the true answer.

We will simply plug our algorithm into the Transfer Theorem, as stated below.
\begin{thm}[Transfer Theorem \cite{BassilyNSSSU16}] \label{thm:transfer}
Let $\mathcal M$ be an interactive algorithm that takes a dataset $X \in \mathcal U^n$ and answers adaptively chosen statistical queries.  Suppose that $\mathcal M$ is 
\begin{enumerate}
\item $(\alpha, \alpha \beta)$-differentially private with respect to the dataset, and also
\item $(\alpha, \alpha \beta)$-accurate with respect to the dataset for $m$ queries.  
\end{enumerate}
Then $\mathcal M$ is $(O(\alpha), O(\beta))$-accurate with respect to the population for $m$ queries.
\end{thm}
With the right choice of parameter $\alpha$, we have the following corollary:
\begin{corollary}
    For any $\beta \in (0,1)$ and $c>0$, there exists an interactive algorithm $\mathcal M$ that takes a dataset $X\in \mathcal{U}^n$, answers an adaptively chosen $m\times N$ statistical query matrix $Q$, and is \[\left(O\!\left(
\sqrt\frac{\gamma}{n}
\log^{5/4}\frac{Nmn}{\beta}\right), \beta
\right)\text{-accurate},\]
where 
 \[\gamma
    =
    \Theta\!\left(
        \gamma_2(Q)\,
            \log^{\frac12+c}\gamma_2(Q)
        \log^3 N
    \right).\]
\end{corollary}
Comparing with \cite{BassilyNSSSU16}, we improve the error from $\tilde{O}(\frac{\sqrt[4]{m}}{\sqrt n})$ to $\tilde{O}(\sqrt\frac{\gamma_2(Q)}{n})$, since $\sqrt{m}$ is an upper bound for the $\gamma_2$ norm of any $m$ statistical queries.

\section*{Acknowledgments and AI Use Statement}

We started studying the adaptive setting after finishing the paper~\cite{NTU26}. Originally, we believed there is a polynomial separation between error bounds achievable in the oblivious and adaptive query settings. We studied a particular lower bound construction in order to show such a separation. While working on a lower bound with GPT6-Astra, the model proposed an algorithm for our lower bound construction, showing that the lower bound we conjectured doesn't  hold. We then used GPT6-Astra to extend the algorithm sufficiently so that it can be combined with the online factorization from~\cite{NTU26}, and give the main upper bound in this paper. With the help of GPT6-Astra, we simplified the algorithm significantly, connected it with the IDC framework, and reworked the privacy analysis so that it fits within (an extension of) the target charging framework. The two main ideas suggested by the LLM were to use a layered algorithm, and when updating the public point, to use a one-shot update to match exactly to the noisy estimate.

The main ideas of the lower bound proof are ours, but the arguments were simplified and streamlined with the help of GPT6-Astra. In particular, the new version of Lemma~\ref{lem:bdkt}, which can be found in \hyperlink{app:lem:bdkt}{Appendix~\ref*{app:proofs}}, is due to GPT6-Astra, whereas our original argument used a modification of the proof in~\cite{BDKT12}. The proof of Lemma~\ref{lem:lb_base} was also given by the same model. We originally had a more combinatorial but significantly more complicated proof of the lower  bound for the same choice of query sequences. 

We would like to thank Roodabeh Safavi for thoughtful discussion on understanding the online version of \cite[Theorem~4.3]{BDKT12}.
\haohua{Also should we thank the discussion with Roodabeh Safavi at ISTA? We looked at bdkt together}\sasho{I think we should do this for the arxiv/proceedings version, not the submission.}


\bibliography{biblio}
\bibliographystyle{alpha}
\appendix

\section{Additional Proofs}
\label{app:proofs}
To prove Proposition~\ref{prop:SDP}, let us first introduce the following technical lemma.
\begin{lemma}[\cite{NTU26}, Lemma~3.3]\label{lm:fact-psd}
    For any $m\times N$ matrix $Q$, and $m\times k$ matrix $L$, there exists a factorization $Q = LR$ with $\|R\|_{1\to 2} \le 1$ if and only if $LL^T \succeq q q^T$ for every column $q$ of $Q$. Moreover, if this condition holds, then we can take $R = MQ$, where $M$ is any $k\times m$ matrix satisfying $LML = L$ and $MLL^TM^T \preceq I$. 
\end{lemma} 
\begin{sdpappendix}
\growingMatrixStatement
\end{sdpappendix}
\begin{proof}[Proof of Proposition~\ref{prop:SDP}]
    The converse is easy to see by setting $\Sigma_t = L_t L_t^T$ and using Lemma~\ref{lm:fact-psd}. We only prove the forward direction below.

    Using induction on $t$, we construct two sequences of matrices $L_1, \ldots, L_m$, and $M_1, \ldots, M_m$, so that
    \begin{itemize}
        \item $L_t$ and $M_t$ are $t\times t$ lower triangular matrices that only depend on $\Sigma_1, \ldots, \Sigma_t$;
        \item $L_t$ is the principal submatrix of $L_{t+1}$ consisting of its top $t$ rows and columns, and $M_t$ is the principal submatrix of $M_{t+1}$ consisting of its top $t$ rows and columns;
        \item $L_tL_t^T = \Sigma_t$.
        \item $L_t M_t L_t = L_t$;
        \item $M_t L_t$ is a diagonal matrix with diagonal entries in $\{0,1\}$.
    \end{itemize}
    We then set $R_t = M_t Q_t$. It follows from the properties above, and Lemma~\ref{lm:fact-psd} that this defines an online factorization mechanism with factorization value $\gamma$. 

    In the base case $t=1$, $\Sigma_1$ is just a non-negative number $\sigma$ such that $\sigma \ge \max_j q_{1,j}^2$ for every entry $q_{1,j}$ of the first row $q_1$. We set $L_1 = (\sqrt{\sigma})$ and $M_1 = ((\sqrt{\sigma})^+)$. Here, and below, we use the notation $c^+$ for
    \[
    c^+ = \begin{cases}
        \frac1c & c\neq 0\\
        0 & c = 0
    \end{cases}.
    \]

    Next we handle the inductive step. Suppose that $\Sigma_{t+1}$ is written as
    \[
    \Sigma_{t+1} = \begin{pmatrix}
        \Sigma_t & x\\
        x^T & \sigma
    \end{pmatrix},
    \]
    for a $t$-dimensional column vector $x$, and a scalar $\sigma\ge 0$.
    It is a standard fact that, since $\Sigma_{t+1}$ is PSD, $x$ is in the range of $\Sigma_t$, and $\sigma \ge x^T \Sigma_t^+ x$, where $\Sigma_t^+$ is the pseudoinverse of $\Sigma_t$. We also have the factorization
    \[
    \Sigma_{t+1} = \begin{pmatrix}
        I & 0\\
        x^T \Sigma_t^+ & 1
    \end{pmatrix}
    \begin{pmatrix}
        \Sigma_t & 0\\
        0 & c
    \end{pmatrix}
    \begin{pmatrix}
        I & \Sigma_t^+ x\\
        0 & 1
    \end{pmatrix},
    \]
    where $c = \sigma - x^T \Sigma_t^+ x \ge 0$ is the Schur complement of $\Sigma_t$ in $\Sigma_{t+1}$. We set 
    \[
    L_{t+1} = 
    \begin{pmatrix}
        I & 0\\
        x^T \Sigma_t^+ & 1
    \end{pmatrix}
    \begin{pmatrix}
        L_t & 0\\
        0 & \sqrt{c}
    \end{pmatrix}=
    \begin{pmatrix}
        L_t & 0\\
        x^T \Sigma_t^+ L_t & \sqrt{c}
    \end{pmatrix},
    \]
    and
    \[
    M_{t+1} = 
    \begin{pmatrix}
        M_t & 0\\
        0 & (\sqrt{c})^+
    \end{pmatrix}
    \begin{pmatrix}
        I & 0\\
        -x^T \Sigma_t^+ & 1
    \end{pmatrix}
    = 
    \begin{pmatrix}
        M_t & 0 \\
        -(\sqrt{c})^+ x^T \Sigma_t^+ & (\sqrt{c})^+
    \end{pmatrix}.
    \]
    The first two properties are obvious. It is also clear from the factorization of $\Sigma_{t+1}$ and the inductive hypothesis that $L_{t+1} L_{t+1}^T = \Sigma_{t+1}$. Let us then verify the last two properties. We have
    \[
    M_{t+1}L_{t+1} = 
    \begin{pmatrix}
        M_t L_t & 0\\
        0 & \mathbf1\{c \neq 0\}
    \end{pmatrix}.
    \]
    By the inductive hypothesis, this is a diagonal matrix with diagonal entries in $\{0,1\}$. Moreover,
    \[
    L_{t+1} M_{t+1}L_{t+1}
    = 
    \begin{pmatrix}
        I & 0\\
        x^T \Sigma_t^+ & 1
    \end{pmatrix}
    \begin{pmatrix}
        L_t M_t L_t & 0\\
        0 & \sqrt{c}
    \end{pmatrix}
    = 
    L_{t+1},
    \]
    where we used the inductive hypothesis to guarantee $L_t M_t L_t = L_t$. This completes the proof. 
\end{proof}

\begin{bdktappendix}
\bdktStatement
\end{bdktappendix}
\begin{proof}[Proof of Lemma~\ref{lem:bdkt}]
    Let us first state the online version of the reduction, Algorithm~\ref{alg:online-bdkt}. The algorithm is almost unchanged compared to~\cite{BDKT12,EdmondsNU20} - it samples a noise vector as the histogram, runs the data-dependent algorithm on it, and applies the resulting error vector to the true answer as additive noise. Note that any valid histogram, that is, a non-negative integer vector, corresponds to a valid dataset. 
\begin{algorithm}[H]
\caption{Online version of BDKT reduction}
\label{alg:online-bdkt}
\begin{algorithmic}[1]
\Require Private histogram $h\in\mathbb{N}^N_{\ge 0}$, $c>0$, and interactive mechanism $\mathcal{M}$.

\State Sample $Z\sim\operatorname{Uniform}\left(\left[ \ceil{1/c} \right]^N\right)$.
\State Initialize one instance of $\mathcal{M}$, using $Z$ as the histogram.

\For{$t=1,\ldots,m$}
    \State Receive query $q_t\in\mathbb{R}^N$.
    \State Input $q_t$ to $\mathcal{M}$ and obtain answer $a_t$.
    \State Output $b_t=\langle q_t,h\rangle
        +a_t-\langle q_t,Z\rangle$.
\EndFor
\end{algorithmic}
\end{algorithm}
We first prove the privacy. To do so, we view the algorithm in a different way. Let $\hat h\coloneq h-Z$. Then \[Z=h-\hat h,\qquad b_t=\iprod{q_t}{\hat h}+a_t.\]
Consider an algorithm that first releases $\hat h$, then runs $\mathcal{M}$ on $h-\hat h$ to obtain the answers $a_t$. We can then compute answers $b_t$ that are identically distributed to the output of Algorithm~\ref{alg:online-bdkt} via post-processing: $b_t = \iprod{q_t}{\hat{h}} + a_t$. The privacy claim now follows by composition, since the release of $\hat h$ is $(0, c)$-differentially private.

Clearly, Algorithm~\ref{alg:online-bdkt} achieves the same error guarantee as $\mathcal{M}$ when the queries are oblivious. Since $Z$ could correspond to a dataset of arbitrary size, we need to take the maximum over all possible dataset sizes. The error bound then clearly follows.
\end{proof}

\begin{weightedappendix}
\weightedTargetChargingStatement
\end{weightedappendix}
\begin{proof}[Proof of Lemma~\ref{lem:weighted_target_charging}]
Our proof will be almost identical to the proof of Theorem B.4 in \cite{CohenL23}, but with an updated counting argument that replaces Equation (1) in the original proof. The next claim captures this argument. 
\begin{claim}[Privacy Cost Counting]\label{clm:privacy_cost_counting}
    Consider a sequence of calls to randomized algorithms $\mathcal{M}_1, \mathcal{M}_2, \ldots: \mathcal{X} \to \mathcal{Y}$, and parameters $\eps_1,\eps_2,...$, s.t. $\forall i,\eps_ i\leq \bar\eps$. Algorithm $\mathcal{M}_i$ and the parameter $\eps_i$ may depend on the outputs of $\mathcal{M}_1, \ldots, \mathcal{M}_{i-1}$. Suppose there is a set $\top_i \subseteq \mathcal{Y}$ (also possibly depending on the outputs of $\mathcal{M}_1, \ldots, \mathcal{M}_{i-1}$) such that, for any input $x \in \mathcal{X}$, conditional on any possible sequence of outputs of $\mathcal{M}_1, \ldots, \mathcal{M}_{i-1}$, $\mathcal{M}_i(x) \in \top_i$ with probability at least $q$. Let $B_i$ be the indicator variable for $\mathcal{M}_i(x) \in \top_i$. If on every execution
    \[
        \sum_i{\eps_i^2B_i} \leq S,
    \]
    then for every $\alpha>0$
    \[
        \Pr\!\left[
        \sum_i\eps_i^2>(1+\alpha)\frac Sq
        \right]
        \le
        \exp\!\left(
        \left(
        \log{(1+\alpha)}-\alpha
        \right)
        \frac{S}{\bar\eps^2}
        \right).
    \]
\end{claim}
\begin{proof}
Let the random variable $\pi_i$ denote the sequence of outputs of $\mathcal{M}_1, \ldots, \mathcal{M}_{i-1}$, with $\pi_0$ denoting the empty sequence.
Observe that, conditioning on calls before the $i$-th call, we have
    \[
          \E[e^{-\lambda\eps_i^2B_i} \mid \pi_i] \le 1-q(1-e^{-\lambda\eps_i^2})\leq \exp\!{
          \left(-\frac{q\eps_i^2(1-e^{-\lambda\bar\eps^2})}{\bar\eps^2}
          \right)}.
    \]
Taking $\lambda=\frac{\log{(1+\alpha)}}{\bar\eps^2}$, and multiplying the above bound over $i$, we obtain
\[
    \E\left[\exp\!\left(
    \frac{q
    (1-\frac{1}{1+\alpha})
    \sum_i\eps_i^2-\log{(1+\alpha)}\sum_i\eps_i^2B_i}{\bar\eps^2}
    \right)\right]\le1.
\]
Since $\sum_i{\eps_i^2B_i} \leq S$, if $\sum_i\eps_i^2>(1+\alpha)\frac Sq$, then we have
\[
    \frac{q
    (1-\frac{1}{1+\alpha})
    \sum_i\eps_i^2-\log{(1+\alpha)}\sum_i\eps_i^2B_i}{\bar\eps^2}\geq \frac{S[\alpha-\log(1+\alpha)]}{\bar{\eps}^2},
\]
and the claim follows from Markov's Inequality.
\end{proof}
We now show what modification needs to be made to the original proof. Similarly to the original proof, we have the same simulator, but we track the accumulated squared privacy cost instead of the number of target hits, and terminate the simulator if this exceeds $\frac{(1+\alpha)S}{q}$. By Claim~\ref{clm:privacy_cost_counting}, this early stop can happen with probability no more than $\exp\left(\!
\bigl(\log(1+\alpha)-\alpha\bigr)
\frac{S}{\bar\varepsilon^2}
\right)$.
Applying fully adaptive composition (\cite[Theorem~2]{WRRW23}) finishes the proof.
\end{proof}

\end{document}